\documentclass[aps,prl,twocolumn,nofootinbib,floatfix,superscriptaddress ]{revtex4-2}

\usepackage{graphicx}
\usepackage{dcolumn}
\usepackage{bm}

\usepackage[english]{babel}

\usepackage{multirow}
\usepackage{amssymb}
\usepackage{amsmath,bbm}
\usepackage[figuresright]{rotating}
\usepackage{capt-of}
\usepackage{xcolor}
\usepackage{booktabs}
\usepackage{siunitx}
\usepackage{placeins}
\usepackage[hidelinks]{hyperref}

\usepackage[english]{babel}
\usepackage[autolanguage]{numprint}

\begin{document}
\title{Lattice calculation of the Sn isotopes near the proton dripline}

\author{Fabian Hildenbrand}
\email{f.hildenbrand@fz-juelich.de}
\affiliation{Institute for Advanced Simulation (IAS-4), Forschungszentrum J\"{u}lich, D-52425 J\"{u}lich, Germany}

\author{Serdar Elhatisari}
\email{selhatisari@gmail.com}
\affiliation{King Fahd University of Petroleum and Minerals (KFUPM), 31261 Dhahran, Saudi Arabia}
\affiliation{Faculty of Natural Sciences and Engineering, Gaziantep Islam Science and Technology University, Gaziantep 27010, Turkey}

\author{Ulf-G.~Mei{\ss}ner}
\email{meissner@hiskp.uni-bonn.de}
\affiliation{Helmholtz-Institut f\"{u}r Strahlen- und Kernphysik and Bethe Center for Theoretical Physics,\\ Universit\"{a}t Bonn, D-53115 Bonn, Germany}
\affiliation{Institute for Advanced Simulation (IAS-4), Forschungszentrum J\"{u}lich, D-52425 J\"{u}lich, Germany}
\affiliation{Peng Huanwu Collaborative Center for Research and Education, International Institute for Interdisciplinary and Frontiers, Beihang University, Beijing 100191, China}

\author{Helen Meyer}
\affiliation{Helmholtz-Institut f\"{u}r Strahlen- und Kernphysik and Bethe Center for Theoretical Physics,\\ Universit\"{a}t Bonn, D-53115 Bonn, Germany}

\author{Zhengxue Ren}
\affiliation{School of Physics, Nankai University, Tianjin 300071, China}

\author{Andreas Herten}
\affiliation{J\"ulich Supercomputing Centre, Forschungszentrum J\"{u}lich, D-52425 J\"{u}lich, Germany}

\author{Mathis Bode}
\affiliation{J\"ulich Supercomputing Centre, Forschungszentrum J\"{u}lich, D-52425 J\"{u}lich, Germany}

\date{\today}

\begin{abstract}
We present the first {\em ab initio} lattice calculations of the proton-rich tin isotopes $^{99}$Sn to $^{102}$Sn using nuclear lattice effective field theory with high-fidelity two- and three-nucleon forces. For a given set of three-nucleon couplings, we reproduce binding energies with $\sim$1\% accuracy for the even--even systems, and obtain energy splitting and two-nucleon separation energies in agreement with experiment. Our results confirm the $N=50$ shell closure and reveal that the binding energy of $^{99}$Sn lies below values extrapolated from heavier isotopes. 
\end{abstract}


\maketitle
\section{Introduction}
The $^{100}$Sn isotope is a particularly fascinating nucleus. It is a rare example of a doubly magic nucleus~\cite{Karthein:2023ric} with equal proton and neutron numbers, $Z=N=50$, and it sits close to the proton dripline of the extensive tin isotopic chain. It also exhibits among the largest known allowed $\beta$-decay strengths and features enhanced $\alpha$ decays~\cite{Clark:2020bum}. These features make it central to both studies of nuclear structure and astrophysical nucleosynthesis. For a detailed overview of the experimental and theoretical developments surrounding this nucleus, we refer the reader to Ref.~\cite{Faestermann:2013ng}. The neighboring isotopes $^{99}$Sn, $^{101}$Sn, and $^{102}$Sn are equally important and pose formidable challenges to {\em ab initio} nuclear theory~\cite{Mougeot:2021hvr}, where weak binding, continuum coupling, and Coulomb effects amplify the impact of many-body truncations and the calibration of three-nucleon (3N) forces. Few regions of the nuclear chart combine such a sharp theoretical challenge with equally rapid experimental progress. Recent advancements at radioactive-ion-beam facilities have enabled high-precision mass measurements, decay spectroscopy, and spectroscopic studies of $^{100}$Sn and nearby isotopes~\cite{Nies:2024gim,Ireland:2024azu}, offering stringent benchmarks for modern nuclear interactions. These developments highlight the urgent need for theoretical frameworks that can treat medium-to-heavy exotic nuclei from first principles. On the theoretical side, significant progress has been made in recent years toward extending {\em ab initio} calculations into the medium-mass and heavy-mass region. A first breakthrough in this direction was achieved in Ref.~\cite{Morris:2017vxi}, in which continuum chiral two- and three-nucleon forces were combined with coupled-cluster and valence-space in-medium similarity renormalization group (VS-IMSRG) methods, yielding first-principles predictions of the doubly magic character of $^{100}$Sn and quadrupole collectivity consistent with a closed shell. More recently, Bogoliubov coupled-cluster theory was used to work out the properties of open-shell Ca, Ni, and Sn isotopes~\cite{Tichai:2023epe}, extending the mass range to $A\simeq180$ and predicting the location of the pertinent neutron driplines. Similarly, the systematics of the tin isotopic chain from $^{100}$Sn to $^{128}$Sn were explored in Ref.~\cite{Scalesi:2024nao}, calculating two-neutron separation energies, the two-neutron shell gap, and the intrinsic axial quadrupole deformation. Note that many of these calculations have relied heavily on the EM~1.8/2.0 chiral interaction~\cite{Hebeler:2010xb}, which combines a precision next-to-next-to-next-to-leading-order (N$^3$LO) two-nucleon force with next-to-next-to-leading-order (N$^2$LO) three-nucleon terms. This interaction describes most features of even--even (and other) nuclei quite well (such as ground-state energies) and is widely used in {\em ab initio} nuclear theory; see, e.g., Refs.~\cite{Stroberg:2019bch,Miyagi:2021pdc,PhysRevLett.117.172501,PhysRevLett.127.242502,Arthuis:2024mnl}. However, it does not resolve the well-known radius problem~\cite{Cipollone:2014hfa,Lapoux:2016exf,LENPIC:2022cyu,Miyagi:2025lmv} and also shows a pronounced regulator (cutoff) dependence. For a review of {\em ab initio} many-body methods combined with chiral forces like EM~1.8/2.0 across the nuclear landscape, see Ref.~\cite{Hergert:2020bxy}. More recently, this interaction was also used to elucidate the structure of the doubly magic nuclei $^{208}$Pb and $^{266}$Pb~\cite{Bonaiti:2025bsb}.

An entirely different approach to nuclear structure and reaction physics is provided by Nuclear Lattice Effective Field Theory (NLEFT). For an introduction, see Ref.~\cite{Lahde:2019npb}, and for a recent review see Ref.~\cite{Lee:2025req}. In NLEFT, space–time is represented by a four-dimensional grid with volume $L^3\times L_t$, where $L$ ($L_t$) is the spatial (temporal) extent, and $L$ is chosen large enough to suppress finite-volume effects. The spatial discretization defines the lattice spacing $a$; here we work with $a=1.32$~fm. This corresponds to a maximum momentum of about $470$~MeV, sufficient to employ chiral nuclear forces. NLEFT calculations with high-fidelity chiral forces at N$^3$LO, combined with the method of wavefunction matching and smeared three-nucleon forces~\cite{Elhatisari:2022zrb}, have already achieved precise results for binding energies and radii of nuclei up to $A=58$ as well as the equation of state of nuclear and neutron matter up to twice nuclear matter density.
With the advent of exascale computational capabilities now available with JUPITER at Forschungszentrum J\"ulich~\cite{jureap}, it has become feasible to extend these investigations into the region of nuclei with $A\simeq100$. A first NLEFT study of $^{100}$Sn using SU(4)-symmetric forces supplemented with a spin–orbit interaction was presented in Ref.~\cite{Niu:2025uxk}.

In this work, we focus on the four tin isotopes $^{99}$Sn--$^{102}$Sn, located near the proton dripline. This study constitutes the first {\em ab initio} lattice investigation in this mass region and extends earlier NLEFT calculations of lighter proton-rich systems, such as $^{22}$Si~\cite{Zhang:2024wfd}, to much heavier nuclei using high-fidelity interactions. Our results should be regarded as a first exploratory but important step into a previously inaccessible domain for NLEFT, providing new insight into the role of three-nucleon forces, shell closures, and proton-rich binding. At the same time, they establish a foundation for systematic lattice studies of the entire tin isotopic chain and its excitations.

\section{Methodology}
We perform lattice calculations of the ground-state energies of the proton-rich isotopes $^{99\text{--}102}$Sn closest to the assumed proton dripline.  We employ NLEFT with high-fidelity chiral interactions at N$^3$LO, as developed in Ref.~\cite{Elhatisari:2022zrb}. The Hamiltonian includes non-locally regulated two-nucleon (2N) forces, where the appearing low-energy constants (LECs) are fitted to neutron-proton scattering data, as well as three-nucleon (3N) interactions consisting of locally smeared contact terms, one-pion exchange terms with locally smeared two-nucleon contacts, and the two-pion exchange potential, whose LECs are fixed from pion-nucleon 
scattering~\cite{Hoferichter:2015tha}. In addition, two SU(4) symmetric terms, $V_{c_E}^{(l)}$ and $V_{c_E}^{(t)}$, are included, yielding a total of eight independent LECs constrained by the ground state
energies of light- and medium mass nuclei up to $^{40}$Ca, and the same set is employed in this work without further adjustment. The Hamiltonian also includes the Coulomb interaction, which in the counting applied here is a next-to-leading order effect, and Galilean invariance restoration (GIR) terms that restore the proper dispersion relations affected by non-local smearing interactions~\cite{Li:2019ldq}.

We employ Euclidean time projection to extract ground-state properties. During the Euclidean time evolution, we use a simplified Hamiltonian $H_s$ consisting of a regulated one-pion exchange as described in Ref.~\cite{Reinert:2017usi} and an SU(4) symmetric two-nucleon short-range  interaction. The difference between $H_s$ and the full chiral Hamiltonian $H$ is treated perturbatively via wavefunction matching method as detailed in Ref.~\cite{Elhatisari:2022zrb}. This approach allows us to perform calculations using high-fidelity interactions with a reduced sign-problem. 

Simulations are carried out in a periodic cubic volume of $L=12$ lattice units (l.u.) with spatial lattice spacing $a=1.32$~fm, which corresponds to a box length of $15.84$~fm. Note that this length is surpassing the phenomenological radius of $^{100}$Sn, \mbox{$R=1.3~{\rm fm}\times A^{1/3}\sim 6$~fm}, by more than a factor of two, thereby minimizing finite volume effects. Furthermore, the recent calculations in Ref.~\cite{Niu:2025uxk}, which were mostly performed at $L=11$~l.u., confirm that this choice is sufficient.
We also note that the chosen lattice spacing corresponds to a maximum momentum of \qty{470}{\mega\electronvolt}, which is a preferred momentum scale in view of the
constraints from large-$N_c$ in QCD~\cite{Lee:2020esp}.
\begin{figure}[tb]
  \begin{center}
          \includegraphics[width=\linewidth]{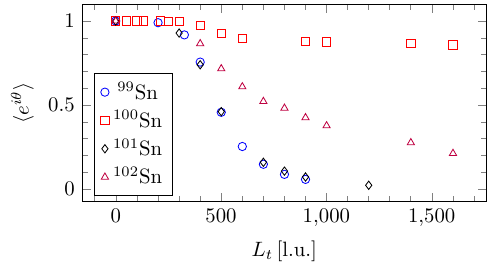}
    \caption{The average phase $\langle e^{i\theta}\rangle$ as a function of time steps $L_t$.}
    \label{fig:phase}
  \end{center}

\end{figure}

For simulations of even--even nuclei at $N=Z$, the NLEFT sign problem is strongly suppressed, allowing calculations with reliable statistics, in particular for three-body operators. The impact for even--even nuclei close to neutron-proton equality $N\simeq Z$ is moderate. We perform such calculations starting from shell-model initial states for up to $L_t=1600$ time steps, using the standard temporal lattice spacing $a_t  =(1000\,{\rm MeV})^{-1}$. For the even-odd nuclei $^{99}$Sn and $^{101}$Sn, reasonable calculations are limited to  $L_t\simeq1000$. The dependence of the corresponding average complex phase 
\begin{equation}
\langle e^{i\theta}\rangle =\langle {\rm det}({\cal M})/|{\rm det}({\cal M})|\rangle~,
\end{equation}
with ${\cal M}$ the transfer matrix is depicted in Fig.~\ref{fig:phase}. Not unexpectedly, the phase of the even--odd nuclei is decreasing significantly faster than the one for even--even nuclei, yet there is still a sufficiently
strong signal to precisely extract the corresponding ground state energies of $^{99}$Sn and $^{101}$Sn. We also note that the required Euclidean time evolution for high-fidelity interactions exceeds the ones needed for a pure SU(4) symmetric calculation substantially, compare with Ref.~\cite{Niu:2025uxk}.

As detailed in Ref.~\cite{Elhatisari:2022zrb} the SU(4)-symmetric 2N coupling $C_s$ is optimized by minimizing the ground-state energy. Since calculations here are computationally high demanding, we perform this analysis only for $^{100}$Sn at $L_t = 500$, where the phase $\langle e^{i\theta}\rangle$ is well under control. The resulting value, $C_s = 0.42 \times 10^{-6}$ MeV$^{-2}$, is then held fixed for all subsequent simulations. The total amount of supercomputer time used on JUPITER is $\sim 36\times10^6$ core hours corresponding to about $\sim 500\times10^3$ GPU hours.

Expectation values of operators, including sums of 2N terms at different orders and all 3N operators, are extrapolated from finite Euclidean times to $\tau\to\infty$ using a combined extrapolation approach with a double-exponential ansatz, accounting for the second order correction of the wave function,
\begin{equation}
    \langle \mathcal{O}(\tau)\rangle=\langle\mathcal{O}(\infty)\rangle+a \exp\left(-\Delta E\tau \right)
    +b \exp\left(-{\Delta E\tau}/{2}\right)\,,
\end{equation}
where  $\tau$ is the Euclidean time, $\mathcal{O}(\infty)$ corresponds to the asymptotic value, and $a,b$ and $\Delta E$ are fit parameters. The decay parameter $\Delta E$ is shared among operators involved in the fit.
As in Ref.~\cite{Elhatisari:2022zrb}, we fit the different 2N and 3N  operator structures separately and then combine these to the final result.
Although not all operators are evolved to maximal $L_t$ due to the computational costs and the convergence behaviour (see Fig.~\ref{fig:phase}), the double exponential form provides a good description for all operators. 

\FloatBarrier
\section{Results and discussion}

\begin{figure}[htb]
  \begin{center}
          \includegraphics[width=\linewidth]{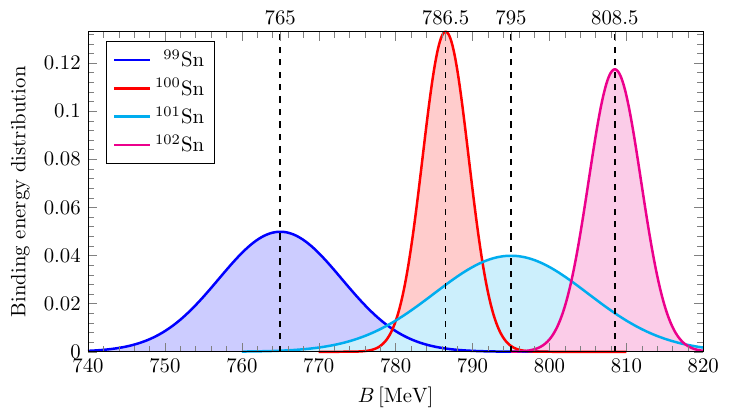}
    \caption{Distribution of the obtained binding energies $B$ for $^{99}$Sn -- $^{102}$Sn. The central values are indicated by the dashed black lines and given on the top axis of the plot. For further details, see Tab.~\ref{tab:energies}.}
    \label{fig:B}
  \end{center}
\end{figure}

The distributions of the ground-state energies of the isotopes $^{99-102}$Sn obtained with NLEFT at N$^3$LO are shown in \autoref{fig:B}, with central values and uncertainties obtained from correlated extrapolations of the individual operator contributions. 
The results, together with the modified 3N interaction (N$^3$LO*), are summarized in \autoref{tab:energies}, alongside experimental values.
\renewcommand{\arraystretch}{1.1}
\begin{table}[htb]
    \sisetup{table-alignment=center, separate-uncertainty, table-align-uncertainty}
    \begin{center}
    \caption{Ground state binding energy of selected Sn isotopes at N$^3$LO with the LECs from Ref.~\cite{Elhatisari:2022zrb}  (N$^3$LO)
    and two modified 3N LECs as described in the text (N$^3$LO*). The experimental results are taken from  Ref.~\cite{Wang:2021xhn}. Note that the experimental value for $^{99}$Sn is extrapolated.  All Energies are given in MeV. \\ }
    \begin{tabular}{c@{\quad}S[table-format=3.1+-2.1]@{\quad\ }S[table-format=3.1+-2.1]@{\quad\ }S[table-format=3.2+-1.2]}
        \toprule 
        {Nucleus} & {N$^3$LO~\cite{Elhatisari:2022zrb}} & {N$^3$LO*} &{Exp.} \\
         \midrule
         {$^{\phantom{1}99}$Sn} &765\pm 8& 804\pm 8 & 807.9\pm 0.6  \\
         {$^{100}$Sn} & 786.5\pm3.0 & 825.2\pm3.0&825.16\pm0.24  \\
         {$^{101}$Sn} &795\pm10 & 834\pm10 & 836.39\pm0.30  \\
         {$^{102}$Sn} &808.5\pm3.4& 848.1\pm3.4 &849.09\pm0.10\\
         \bottomrule
    \end{tabular}
    \label{tab:energies}
    \end{center}
\end{table}
We note that the precision of the even–even isotopes is visibly superior to that of the odd-$A$ systems, which is consistent with expectations from Fig.~\ref{fig:phase}. This reflects the suppression of the sign problem for $N=Z$ nuclei, which allows for tighter constraints on three-body operators at long Euclidean times. For odd-$A$ isotopes, statistical degradation manifests in the larger uncertainties quoted in \autoref{tab:energies}. For all isotopes, calculations with the original N$^3$LO Hamiltonian systematically underbind by about 5\% relative to experiment~\cite{Wang:2021xhn}. This is not unexpected, as the three-nucleon forces were originally constrained in the light-to-medium mass region up to $^{40}$Ca, some sixty nucleons lighter than the systems considered here~\cite{Elhatisari:2022zrb}. 

To assess the role of three-nucleon forces, we performed controlled variations of two SU(4)-symmetric 3N couplings, $V_{c_E}^{(l)}$ and $V_{c_E}^{(t)}$, originally motivated by $\alpha$-cluster EFT. With only a percent-level adjustment, we reproduce the experimental binding energy of $^{100}$Sn (denoted N$^3$LO* in \autoref{tab:energies}). Remarkably, this modification brings the neighboring isotopes $^{99}$Sn, $^{101}$Sn, and $^{102}$Sn into quantitative agreement with experiment without further tuning. Such behavior points to a systematic underestimation of the effective 3N contribution in heavier nuclei, plausibly reflecting less prominent relevance of $\alpha$-cluster–motivated operators in heavier nuclei, see e.g.~\cite{Delion:2023qha}. Importantly, the required modifications in LECs remain small enough to preserve good agreement with the nuclei originally used in the fit, such as shifts of $1.7\%$ for $^{40}$Ca and $0.7\%$ for $^4$He.

Mass differences between the isotopes provide an even more stringent benchmark, and we extract the two-neutron separation energy $S_{2n}$,
\begin{align}\label{eq:2n}
    S_{2n} = E(Z,N)-E(Z,N-2)\,,
\end{align}
with $E(Z,N)$ the energy of the nucleus with the proton number $Z$ and the neutron number $N$. 
\renewcommand{\arraystretch}{1.1}
\begin{table}[htb]
    \sisetup{table-alignment=center, separate-uncertainty, table-align-uncertainty, table-format=2.1+-2.1}
    \begin{center}
    \caption{Mass splittings. The $^\star$ denotes a two-neutron separation energy $S_{2n}$, see Eq.~\eqref{eq:2n}.
    All energies are given in MeV. For further details, see Tab.~\ref{tab:energies}.\\}
    \begin{tabular}{c@{\quad}S@{\quad\ }S@{\quad\ }S}
        \toprule
        {Isotopes} & {N$^3$LO} & {N$^3$LO*} & {Exp.} \\
         \midrule\midrule
         {102-101} &  13.5\pm 10.6 & 14.1 \pm 10.6 & 12.7 \pm 0.3\\
         {102-100$^\star$} &  22.5 \pm 4.6  & 22.9 \pm 4.6  & 23.9 \pm 0.3\\
         {102-99}  &  43.5 \pm 8.7  & 44.1 \pm 8.7  & 41.2 \pm 0.6 \\
         \midrule
         {101-100} &  8.5 \pm 10.4  & 8.9 \pm 10.4 & 11.2 \pm 0.4\\
         {101-99$^\star$}  &  30.0 \pm 12.8 & 30.1 \pm 12.8 & 28.5 \pm 0.7\\
         \midrule
         {100-99}   & 21.5 \pm 8.6 & 21.2 \pm 8.6  &  17.3 \pm 0.7 \\
         \bottomrule
    \end{tabular}
    \label{tab:splitt}
    \end{center}
\end{table}
The results are presented in \autoref{tab:splitt} and we observe overall consistency with other theoretical results, see e.g. Ref.~\cite{Tichai:2023epe}. \autoref{tab:splitt} shows that the N$^3$LO* Hamiltonian reproduces two-neutron separation energies and nearest-neighbor splittings to within experimental uncertainties. In particular, the calculated $S_{2n}(^{102}$Sn$)=\qty{22.9(4.6)}{\mega\electronvolt}$ compares favorably with the precise experimental value of \qty{23.9(3)}{\mega\electronvolt}. The $^{99}$Sn–$^{100}$Sn splitting shows a \qty{\sim 10}{\mega\electronvolt} enhancement relative to neighboring pairs, consistent with the expected $N=50$ shell closure~\cite{Mayer:1948zz}. Thus, our results provide an ab initio confirmation of the doubly magic nature of $^{100}$Sn, confirming earlier coupled-cluster predictions~\cite{Morris:2017vxi}.

These findings are noteworthy when compared to other state-of-the-art approaches. Coupled-cluster calculations in the tin region~\cite{Morris:2017vxi} and VS-IMSRG studies up to $^{128}$Sn~\cite{Tichai:2023epe} reproduce global separation energy trends, but typically exhibit interaction-dependent discrepancies in absolute binding energies. In contrast, NLEFT with high-fidelity chiral forces and modest 3N adjustments captures both absolute and differential observables in the proton-rich regime.

The physical origin of the successful reproduction of experimental energies can be traced to the delicate interplay of strong 2N, 3N interactions as well as the Coulomb repulsion between protons. For instance, in $^{100}$Sn the kinetic plus two-nucleon interaction contributions to the binding energy amounts to \qty{717.9}{\mega\electronvolt}, balanced by a large repulsive Coulomb contribution of \qty{-366.8}{\mega\electronvolt} and an attractive 3N contribution of \qty{435.6}{\mega\electronvolt}. The net result matches experiment only when the 3N sector is slightly enhanced, underscoring its central role in stabilizing proton-rich systems against Coulomb repulsion. Interestingly, the Coulomb contribution is lower than  a modern version of the modified Bethe-Weizs\"acker formula~\cite{Cao:2022kny}  and differs by \qty{\sim20}{\mega\electronvolt} from earlier SU(4)-based lattice calculations~\cite{Niu:2025uxk}, which is even further away, highlighting the importance of high-fidelity chiral interactions in quantifying fine details.

Finally, experimental benchmarks provide additional perspective. Recent experimental progress has sharpened the benchmarks for ab initio theory in the proton-rich tin region. High-precision mass measurements of $^{103}$Sn have reduced uncertainties by more than an order of magnitude, reestablishing a smooth mass surface near the doubly magic $^{100}$Sn and correcting earlier irregularities in the Atomic Mass Evaluation~\cite{Ireland:2024azu}. Complementary ISOLTRAP measurements have further refined the nuclear mass surface by reducing the uncertainty in the $^{103}$Sn mass by a factor of four, leading to revised and more bound mass values for $^{101}$Sn and neighboring isotopes~\cite{Nies:2024gim}. These developments offer a natural target for future lattice investigations and our present calculations demonstrate, for the first time, that lattice EFT can achieve quantitative agreement with the new mass benchmarks in a mass region previously considered inaccessible for ab initio lattice methods.

\section{Summary and outlook}

We have performed the first {\em ab initio} lattice calculations of the proton-rich tin isotopes $^{99}$Sn--$^{102}$Sn in the framework of NLEFT,
based on high-fidelity chiral forces at N$^3$LO previously used to explore the mass region from the deuteron to $^{58}$Ni~\cite{Elhatisari:2022zrb,Ma:2023ahg,Konig:2023rwe,Hildenbrand:2024ypw,Elhatisari:2024otn,Shen:2024qzi,Zhang:2024wfd,Elhatisari:2025fyu}. The three-nucleon LECs entering these interactions were originally constrained up to $^{40}$Ca~\cite{Elhatisari:2022zrb}, well below the mass region considered here. Using these N$^3$LO forces as is, we observe a systematic underbinding of about $5\%$ in absolute energies, while mass differences and two-neutron separation energies are consistent with experiment within the uncertainties. As expected from the suppressed sign problem at $N=Z$, the uncertainties in the binding energies of the even--even isotopes are below $1\%$, whereas they are of the order of $1\%$ for the odd isotopes.

To assess the role of three-nucleon forces in this heavier, proton-rich regime, we have carried out a controlled, slight retuning of two particular 3NF operators. This percent-level modification leads to an almost exact reproduction of the
experimental values for $^{100}$Sn, $^{101}$Sn and $^{102}$Sn and reduces the central value the $^{99}$Sn binding energy compared to the extrapolated value~\cite{Wang:2021xhn}. Importantly, this adjustment retains the good description of the nuclei reported in the original fit. Therefore, our investigation indicates that a more comprehensive calibration of three-nucleon forces, incorporating data from heavier nuclei, is necessary. This might also require the inclusion of reaction data, as recently pointed out in Ref.~\cite{Elhatisari:2025fyu}. Furthermore, it will be interesting to explore the whole Sn isotope chain and also consider excited states (in particular $2^+$), charge radii,
$\beta$-decays, and collective excitations. Such work is underway but requires substantial computational resources. 

\section{Acknowledgements}

This project ExaNP is part of JUREAP, the JUPITER Research and Early Access Program at the J\"ulich Supercomputing Centre (JSC), and was selected as a GCS Lighthouse project in the Exascale Pioneer Call for JUPITER. We gratefully acknowledge the awarded compute time. 
This work was supported in part by the European
Research Council (ERC) under the European Union's Horizon 2020 research
and innovation programme (grant agreement No. 101018170),
and by the CAS President's International Fellowship Initiative (PIFI) (Grant No.~2025PD0022). The work of SE is supported
in part by the Scientific and Technological Research Council of Turkey (TUBITAK project no. 123F464). The authors gratefully acknowledge the Gauss Centre for Supercomputing e.V. (www.gauss-centre.eu)
for funding this project by providing computing time on the GCS Supercomputer JUWELS
at J\"ulich Supercomputing Centre (JSC) and the support of the project \mbox{EXOTIC} by the JSC by dedicated HPC time provided on the \mbox{JURECA DC} GPU partition, where part of the GPU code was developed.

\bibliographystyle{unsrturl}

\bibliography{references.bib}

\begin{thebibliography}{10}

\bibitem{Karthein:2023ric}
J.~Karthein et~al.
\newblock {Electromagnetic properties of indium isotopes illuminate the doubly
  magic character of $^{100}$Sn}.
\newblock {\em Nature Phys.}, 20(11):1719--1725, 2024.
\newblock \href {https://arxiv.org/abs/2310.15093} {\path{arXiv:2310.15093}},
  \href {https://doi.org/10.1038/s41567-024-02612-y}
  {\path{doi:10.1038/s41567-024-02612-y}}.

\bibitem{Clark:2020bum}
R.~M. Clark, A.~O. Macchiavelli, H.~L. Crawford, P.~Fallon, D.~Rudolph,
  A.~S{\r{a}}mark-Roth, C.~M. Campbell, M.~Cromaz, C.~Morse, and C.~Santamaria.
\newblock {Enhancement of {\ensuremath{\alpha}}-particle formation near
  $^{100}$Sn}.
\newblock {\em Phys. Rev. C}, 101(3):034313, 2020.
\newblock \href {https://doi.org/10.1103/PhysRevC.101.034313}
  {\path{doi:10.1103/PhysRevC.101.034313}}.

\bibitem{Faestermann:2013ng}
T.~Faestermann, M.~Gorska, and H.~Grawe.
\newblock {The structure of Sn-100 and neighbouring nuclei}.
\newblock {\em Prog. Part. Nucl. Phys.}, 69:85--130, 2013.
\newblock \href {https://doi.org/10.1016/j.ppnp.2012.10.002}
  {\path{doi:10.1016/j.ppnp.2012.10.002}}.

\bibitem{Mougeot:2021hvr}
M.~Mougeot et~al.
\newblock {Mass measurements of $^{99-101}$In challenge ab initio nuclear
  theory of the nuclide $^{100}$Sn}.
\newblock {\em Nature Phys.}, 17:1099, 2021.
\newblock \href {https://arxiv.org/abs/2109.10673} {\path{arXiv:2109.10673}},
  \href {https://doi.org/10.1038/s41567-021-01326-9}
  {\path{doi:10.1038/s41567-021-01326-9}}.

\bibitem{Nies:2024gim}
L.~Nies et~al.
\newblock {Refining the nuclear mass surface with the mass of $^{103}$Sn}.
\newblock {\em Phys. Rev. C}, 111(1):014315, 2025.
\newblock \href {https://arxiv.org/abs/2410.17995} {\path{arXiv:2410.17995}},
  \href {https://doi.org/10.1103/PhysRevC.111.014315}
  {\path{doi:10.1103/PhysRevC.111.014315}}.

\bibitem{Ireland:2024azu}
C.~M. Ireland et~al.
\newblock {High-precision mass measurement of $^{103}$Sn restores smoothness of
  the mass surface}.
\newblock {\em Phys. Rev. C}, 111(1):014314, 2025.
\newblock \href {https://arxiv.org/abs/2410.04650} {\path{arXiv:2410.04650}},
  \href {https://doi.org/10.1103/PhysRevC.111.014314}
  {\path{doi:10.1103/PhysRevC.111.014314}}.

\bibitem{Morris:2017vxi}
T.~D. Morris, J.~Simonis, S.~R. Stroberg, C.~Stumpf, G.~Hagen, J.~D. Holt,
  G.~R. Jansen, T.~Papenbrock, R.~Roth, and A.~Schwenk.
\newblock {Structure of the lightest tin isotopes}.
\newblock {\em Phys. Rev. Lett.}, 120(15):152503, 2018.
\newblock \href {https://arxiv.org/abs/1709.02786} {\path{arXiv:1709.02786}},
  \href {https://doi.org/10.1103/PhysRevLett.120.152503}
  {\path{doi:10.1103/PhysRevLett.120.152503}}.

\bibitem{Tichai:2023epe}
A.~Tichai, P.~Demol, and T.~Duguet.
\newblock {Towards heavy-mass ab initio nuclear structure: Open-shell Ca, Ni
  and Sn isotopes from Bogoliubov coupled-cluster theory}.
\newblock {\em Phys. Lett. B}, 851:138571, 2024.
\newblock \href {https://arxiv.org/abs/2307.15619} {\path{arXiv:2307.15619}},
  \href {https://doi.org/10.1016/j.physletb.2024.138571}
  {\path{doi:10.1016/j.physletb.2024.138571}}.

\bibitem{Scalesi:2024nao}
Alberto Scalesi, Thomas Duguet, Pepijn Demol, Mikael Frosini, Vittorio
  Som{\`a}, and Alexander Tichai.
\newblock {Impact of correlations on nuclear binding energies: Ab initio
  calculations of singly and doubly open-shell nuclei}.
\newblock {\em Eur. Phys. J. A}, 60(10):209, 2024.
\newblock \href {https://arxiv.org/abs/2406.03545} {\path{arXiv:2406.03545}},
  \href {https://doi.org/10.1140/epja/s10050-024-01424-1}
  {\path{doi:10.1140/epja/s10050-024-01424-1}}.

\bibitem{Hebeler:2010xb}
K.~Hebeler, S.~K. Bogner, R.~J. Furnstahl, A.~Nogga, and A.~Schwenk.
\newblock {Improved nuclear matter calculations from chiral low-momentum
  interactions}.
\newblock {\em Phys. Rev. C}, 83:031301, 2011.
\newblock \href {https://arxiv.org/abs/1012.3381} {\path{arXiv:1012.3381}},
  \href {https://doi.org/10.1103/PhysRevC.83.031301}
  {\path{doi:10.1103/PhysRevC.83.031301}}.

\bibitem{Stroberg:2019bch}
S.~R. Stroberg, J.~D. Holt, A.~Schwenk, and J.~Simonis.
\newblock {Ab Initio Limits of Atomic Nuclei}.
\newblock {\em Phys. Rev. Lett.}, 126(2):022501, 2021.
\newblock \href {https://arxiv.org/abs/1905.10475} {\path{arXiv:1905.10475}},
  \href {https://doi.org/10.1103/PhysRevLett.126.022501}
  {\path{doi:10.1103/PhysRevLett.126.022501}}.

\bibitem{Miyagi:2021pdc}
T.~Miyagi, S.~R. Stroberg, P.~Navr{\'a}til, K.~Hebeler, and J.~D. Holt.
\newblock {Converged ab initio calculations of heavy nuclei}.
\newblock {\em Phys. Rev. C}, 105(1):014302, 2022.
\newblock \href {https://arxiv.org/abs/2104.04688} {\path{arXiv:2104.04688}},
  \href {https://doi.org/10.1103/PhysRevC.105.014302}
  {\path{doi:10.1103/PhysRevC.105.014302}}.

\bibitem{PhysRevLett.117.172501}
G.~Hagen, G.~R. Jansen, and T.~Papenbrock.
\newblock Structure of $^{78}${Ni} from first-principles computations.
\newblock {\em Phys. Rev. Lett.}, 117:172501, Oct 2016.
\newblock URL: \url{https://link.aps.org/doi/10.1103/PhysRevLett.117.172501},
  \href {https://doi.org/10.1103/PhysRevLett.117.172501}
  {\path{doi:10.1103/PhysRevLett.117.172501}}.

\bibitem{PhysRevLett.127.242502}
R.~Wirth, J.~M. Yao, and H.~Hergert.
\newblock Ab initio calculation of the contact operator contribution in the
  standard mechanism for neutrinoless double beta decay.
\newblock {\em Phys. Rev. Lett.}, 127:242502, Dec 2021.
\newblock URL: \url{https://link.aps.org/doi/10.1103/PhysRevLett.127.242502},
  \href {https://doi.org/10.1103/PhysRevLett.127.242502}
  {\path{doi:10.1103/PhysRevLett.127.242502}}.

\bibitem{Arthuis:2024mnl}
P.~Arthuis, K.~Hebeler, and A.~Schwenk.
\newblock {Neutron-rich nuclei and neutron skins from chiral low-resolution
  interactions}.
\newblock 1 2024.
\newblock \href {https://arxiv.org/abs/2401.06675} {\path{arXiv:2401.06675}}.

\bibitem{Cipollone:2014hfa}
A.~Cipollone, C.~Barbieri, and P.~Navr{\'a}til.
\newblock {Chiral three-nucleon forces and the evolution of correlations along
  the oxygen isotopic chain}.
\newblock {\em Phys. Rev. C}, 92(1):014306, 2015.
\newblock \href {https://arxiv.org/abs/1412.0491} {\path{arXiv:1412.0491}},
  \href {https://doi.org/10.1103/PhysRevC.92.014306}
  {\path{doi:10.1103/PhysRevC.92.014306}}.

\bibitem{Lapoux:2016exf}
V.~Lapoux, V.~Som{\`a}, C.~Barbieri, H.~Hergert, J.~D. Holt, and S.~R.
  Stroberg.
\newblock {Radii and Binding Energies in Oxygen Isotopes: A Challenge for
  Nuclear Forces}.
\newblock {\em Phys. Rev. Lett.}, 117(5):052501, 2016.
\newblock \href {https://arxiv.org/abs/1605.07885} {\path{arXiv:1605.07885}},
  \href {https://doi.org/10.1103/PhysRevLett.117.052501}
  {\path{doi:10.1103/PhysRevLett.117.052501}}.

\bibitem{LENPIC:2022cyu}
P.~Maris et~al.
\newblock {Nuclear properties with semilocal momentum-space regularized chiral
  interactions beyond N2LO}.
\newblock {\em Phys. Rev. C}, 106(6):064002, 2022.
\newblock \href {https://arxiv.org/abs/2206.13303} {\path{arXiv:2206.13303}},
  \href {https://doi.org/10.1103/PhysRevC.106.064002}
  {\path{doi:10.1103/PhysRevC.106.064002}}.

\bibitem{Miyagi:2025lmv}
Takayuki Miyagi.
\newblock {Nuclear radii from first principles}.
\newblock {\em Front. in Phys.}, 13:1581854, 2025.
\newblock \href {https://doi.org/10.3389/fphy.2025.1581854}
  {\path{doi:10.3389/fphy.2025.1581854}}.

\bibitem{Hergert:2020bxy}
H.~Hergert.
\newblock {A Guided Tour of ab initio Nuclear Many-Body Theory}.
\newblock {\em Front. in Phys.}, 8:379, 2020.
\newblock \href {https://arxiv.org/abs/2008.05061} {\path{arXiv:2008.05061}},
  \href {https://doi.org/10.3389/fphy.2020.00379}
  {\path{doi:10.3389/fphy.2020.00379}}.

\bibitem{Bonaiti:2025bsb}
Francesca Bonaiti, Gaute Hagen, and Thomas Papenbrock.
\newblock {Structure of the doubly magic nuclei $^{208}$Pb and $^{266}$Pb from
  ab initio computations}.
\newblock 8 2025.
\newblock \href {https://arxiv.org/abs/2508.14217} {\path{arXiv:2508.14217}}.

\bibitem{Lahde:2019npb}
Timo~A. L{\"a}hde and Ulf-G. Mei{\ss}ner.
\newblock {\em {Nuclear Lattice Effective Field Theory}: {An introduction}},
  volume 957.
\newblock Springer, 2019.
\newblock \href {https://doi.org/10.1007/978-3-030-14189-9}
  {\path{doi:10.1007/978-3-030-14189-9}}.

\bibitem{Lee:2025req}
Dean Lee.
\newblock {Lattice Effective Field Theory Simulations of Nuclei}.
\newblock 1 2025.
\newblock \href {https://arxiv.org/abs/2501.03303} {\path{arXiv:2501.03303}}.

\bibitem{Elhatisari:2022zrb}
Serdar Elhatisari et~al.
\newblock {Wavefunction matching for solving quantum many-body problems}.
\newblock {\em Nature}, 630(8015):59--63, 2024.
\newblock \href {https://arxiv.org/abs/2210.17488} {\path{arXiv:2210.17488}},
  \href {https://doi.org/10.1038/s41586-024-07422-z}
  {\path{doi:10.1038/s41586-024-07422-z}}.

\bibitem{jureap}
J{\"u}lich~Supercomputing Centre.
\newblock Jureap - seeding exascale in europe.
\newblock \url{https://www.fz-juelich.de/en/ias/jsc/jupiter/jureap}, 2025.

\bibitem{Niu:2025uxk}
Zhong-Wang Niu and Bing-Nan Lu.
\newblock {Sign-Problem-Free Nuclear Quantum Monte Carlo}.
\newblock 6 2025.
\newblock \href {https://arxiv.org/abs/2506.12874} {\path{arXiv:2506.12874}}.

\bibitem{Zhang:2024wfd}
Shuang Zhang, Serdar Elhatisari, Ulf-G. Mei{\ss}ner, and Shihang Shen.
\newblock {Lattice simulation of nucleon distribution and shell closure in the
  proton-rich nucleus $^{22}$Si}.
\newblock {\em Phys. Lett. B}, 869:139839, 2025.
\newblock \href {https://arxiv.org/abs/2411.17462} {\path{arXiv:2411.17462}},
  \href {https://doi.org/10.1016/j.physletb.2025.139839}
  {\path{doi:10.1016/j.physletb.2025.139839}}.

\bibitem{Hoferichter:2015tha}
Martin Hoferichter, Jacobo Ruiz~de Elvira, Bastian Kubis, and Ulf-G.
  Mei{\ss}ner.
\newblock {Matching pion-nucleon Roy-Steiner equations to chiral perturbation
  theory}.
\newblock {\em Phys. Rev. Lett.}, 115(19):192301, 2015.
\newblock \href {https://arxiv.org/abs/1507.07552} {\path{arXiv:1507.07552}},
  \href {https://doi.org/10.1103/PhysRevLett.115.192301}
  {\path{doi:10.1103/PhysRevLett.115.192301}}.

\bibitem{Li:2019ldq}
Ning Li, Serdar Elhatisari, Evgeny Epelbaum, Dean Lee, Bingnan Lu, and Ulf-G
  Mei{\ss}ner.
\newblock {Galilean invariance restoration on the lattice}.
\newblock {\em Phys. Rev. C}, 99(6):064001, 2019.
\newblock \href {https://arxiv.org/abs/1902.01295} {\path{arXiv:1902.01295}},
  \href {https://doi.org/10.1103/PhysRevC.99.064001}
  {\path{doi:10.1103/PhysRevC.99.064001}}.

\bibitem{Reinert:2017usi}
P.~Reinert, H.~Krebs, and E.~Epelbaum.
\newblock {Semilocal momentum-space regularized chiral two-nucleon potentials
  up to fifth order}.
\newblock {\em Eur. Phys. J. A}, 54(5):86, 2018.
\newblock \href {https://arxiv.org/abs/1711.08821} {\path{arXiv:1711.08821}},
  \href {https://doi.org/10.1140/epja/i2018-12516-4}
  {\path{doi:10.1140/epja/i2018-12516-4}}.

\bibitem{Lee:2020esp}
Dean Lee et~al.
\newblock {Hidden Spin-Isospin Exchange Symmetry}.
\newblock {\em Phys. Rev. Lett.}, 127(6):062501, 2021.
\newblock \href {https://arxiv.org/abs/2010.09420} {\path{arXiv:2010.09420}},
  \href {https://doi.org/10.1103/PhysRevLett.127.062501}
  {\path{doi:10.1103/PhysRevLett.127.062501}}.

\bibitem{Wang:2021xhn}
Meng Wang, W.~J. Huang, F.~G. Kondev, G.~Audi, and S.~Naimi.
\newblock {The AME 2020 atomic mass evaluation (II). Tables, graphs and
  references}.
\newblock {\em Chin. Phys. C}, 45(3):030003, 2021.
\newblock \href {https://doi.org/10.1088/1674-1137/abddaf}
  {\path{doi:10.1088/1674-1137/abddaf}}.

\bibitem{Delion:2023qha}
D.~S. Delion and A.~Dumitrescu.
\newblock {Alpha-clustering and related phenomena in medium and heavy nuclei}.
\newblock {\em Eur. Phys. J. A}, 59(9):210, 2023.
\newblock \href {https://doi.org/10.1140/epja/s10050-023-01105-5}
  {\path{doi:10.1140/epja/s10050-023-01105-5}}.

\bibitem{Mayer:1948zz}
Maria~G. Mayer.
\newblock {On Closed Shells in Nuclei}.
\newblock {\em Phys. Rev.}, 74:235--239, 1948.
\newblock \href {https://doi.org/10.1103/PhysRev.74.235}
  {\path{doi:10.1103/PhysRev.74.235}}.

\bibitem{Cao:2022kny}
Yuping Cao, Danhui Lu, Yibin Qian, and Zhongzhou Ren.
\newblock {Uncertainty analysis for the nuclear liquid drop model and
  implications for the symmetry energy coefficients}.
\newblock {\em Phys. Rev. C}, 105(3):034304, 2022.
\newblock \href {https://doi.org/10.1103/PhysRevC.105.034304}
  {\path{doi:10.1103/PhysRevC.105.034304}}.

\bibitem{Ma:2023ahg}
Yuan-Zhuo Ma, Zidu Lin, Bing-Nan Lu, Serdar Elhatisari, Dean Lee, Ning Li,
  Ulf-G. Mei{\ss}ner, Andrew~W. Steiner, and Qian Wang.
\newblock {Structure Factors for Hot Neutron Matter from Ab~Initio Lattice
  Simulations with High-Fidelity Chiral Interactions}.
\newblock {\em Phys. Rev. Lett.}, 132(23):232502, 2024.
\newblock \href {https://arxiv.org/abs/2306.04500} {\path{arXiv:2306.04500}},
  \href {https://doi.org/10.1103/PhysRevLett.132.232502}
  {\path{doi:10.1103/PhysRevLett.132.232502}}.

\bibitem{Konig:2023rwe}
Kristian K{\"o}nig et~al.
\newblock {Nuclear Charge Radii of Silicon Isotopes}.
\newblock {\em Phys. Rev. Lett.}, 132(16):162502, 2024.
\newblock [Erratum: Phys.Rev.Lett. 133, 059901 (2024)].
\newblock \href {https://arxiv.org/abs/2309.02037} {\path{arXiv:2309.02037}},
  \href {https://doi.org/10.1103/PhysRevLett.132.162502}
  {\path{doi:10.1103/PhysRevLett.132.162502}}.

\bibitem{Hildenbrand:2024ypw}
Fabian Hildenbrand, Serdar Elhatisari, Zhengxue Ren, and Ulf-G. Mei{\ss}ner.
\newblock {Towards hypernuclei from nuclear lattice effective field theory}.
\newblock {\em Eur. Phys. J. A}, 60(10):215, 2024.
\newblock \href {https://arxiv.org/abs/2406.17638} {\path{arXiv:2406.17638}},
  \href {https://doi.org/10.1140/epja/s10050-024-01427-y}
  {\path{doi:10.1140/epja/s10050-024-01427-y}}.

\bibitem{Elhatisari:2024otn}
Serdar Elhatisari, Fabian Hildenbrand, and Ulf-G. Mei{\ss}ner.
\newblock {The triton lifetime from nuclear lattice effective field theory}.
\newblock {\em Phys. Lett. B}, 859:139086, 2024.
\newblock \href {https://arxiv.org/abs/2408.06670} {\path{arXiv:2408.06670}},
  \href {https://doi.org/10.1016/j.physletb.2024.139086}
  {\path{doi:10.1016/j.physletb.2024.139086}}.

\bibitem{Shen:2024qzi}
Shihang Shen, Serdar Elhatisari, Dean Lee, Ulf-G. Mei{\ss}ner, and Zhengxue
  Ren.
\newblock {Ab~Initio Study of the Beryllium Isotopes Be7 to Be12}.
\newblock {\em Phys. Rev. Lett.}, 134(16):162503, 2025.
\newblock \href {https://arxiv.org/abs/2411.14935} {\path{arXiv:2411.14935}},
  \href {https://doi.org/10.1103/PhysRevLett.134.162503}
  {\path{doi:10.1103/PhysRevLett.134.162503}}.

\bibitem{Elhatisari:2025fyu}
Serdar Elhatisari, Fabian Hildenbrand, and Ulf-G. Mei{\ss}ner.
\newblock {Ab initio lattice study of neutron-alpha scattering with chiral
  forces at N3LO}.
\newblock 7 2025.
\newblock \href {https://arxiv.org/abs/2507.08495} {\path{arXiv:2507.08495}}.

\end{thebibliography}

\end{document}